\documentclass[prd,twocolumn,floatfix,showpacs,superscriptaddress]{revtex4}

\usepackage{graphicx}
\usepackage{epsfig}
\usepackage{bm}
\usepackage{amssymb}
\usepackage{float}
\usepackage{amsmath}
\usepackage{dcolumn}
\usepackage{cancel}
\usepackage[colorlinks]{hyperref}
\usepackage[usenames,dvipsnames]{color}
\hypersetup{
     breaklinks=true,
    pdfstartview={FitH},    
    colorlinks=true,       
    linkcolor=blue,          
    citecolor=red,        
    filecolor=magenta,      
    urlcolor=blue,           
    anchorcolor=green,      
    linktocpage=true
}
\newcommand{\newc}{\newcommand}
\newc{\be}{\begin{equation}}
\newc{\ee}{\end{equation}}

\newc{\ba}{\begin{eqnarray}}
\newc{\ea}{\end{eqnarray}}
\newc{\bea}{\begin{eqnarray*}}
\newc{\eea}{\end{eqnarray*}}
\newc{\D}{\partial}
\newc{\ie}{{\it i.e.} }
\newc{\eg}{{\it e.g.} }
\newc{\etc}{{\it etc.} }
\newc{\etal}{{\it et al.}}
\newc{\lcdm}{$\Lambda$CDM }

\newc{\ra}{\Rightarrow}

\def\doi{http://doi.org}

\begin{document}

\title{Big Bang Nucleosynthesis constraints on Barrow entropy\footnote{This 
work is dedicated to our colleague John D. Barrow, who proposed us the idea and 
worked on the manuscript until its final stage, and passed away on 26 
September 2020.}}

\author{John D. Barrow \footnote{Deceased}}  
\affiliation{DAMTP, Centre for Mathematical Sciences,
University of Cambridge,
Wilberforce Road, Cambridge CB3 0WA
United Kingdom}
 
\author{Spyros Basilakos}\email{svasil@academyofathens.gr}
\affiliation{Academy of Athens, Research Center for Astronomy and
Applied Mathematics, Soranou Efesiou 4, 11527, Athens, Greece}
\affiliation{National Observatory of Athens, Lofos Nymfon, 11852 Athens, 
Greece}

\author{Emmanuel N. Saridakis}
\email{msaridak@noa.gr}
\affiliation{National Observatory of Athens, Lofos Nymfon, 11852 Athens, 
Greece}
\affiliation{CAS Key Laboratory for Researches in Galaxies and Cosmology, 
Department of Astronomy, University of Science and Technology of China, Hefei, 
Anhui 230026, P.R. China}
\affiliation{School of Astronomy, School of Physical Sciences, 
University of Science and Technology of China, Hefei 230026, P.R. China}

\begin{abstract}  
We use Big Bang Nucleosynthesis (BBN) data in order to impose constraints on 
the exponent  of Barrow entropy. The latter is 
an extended entropy relation arising from the incorporation of 
quantum-gravitational effects on the black-hole structure, parameterized 
effectively by the new parameter $\Delta$. When considered in 
a cosmological framework and under the light of the gravity-thermodynamics 
conjecture, Barrow entropy leads to modified cosmological scenarios whose 
Friedmann equations contain extra terms.   We perform a detailed analysis of 
the BBN era and we calculate the deviation of the freeze-out
temperature comparing to the result of standard cosmology. We use the 
observationally determined bound on $ \left|\frac{\delta {T}_f}{{T}_f}\right|$ 
in order to extract the 
upper bound on $\Delta$. As we find, the Barrow exponent should be 
inside the bound  $\Delta\lesssim 1.4\times 10^{-4}$ in order not to spoil the 
BBN epoch, which shows that the deformation from 
standard  Bekenstein-Hawking expression should be small as 
expected.

\end{abstract}


\maketitle

\section{Introduction}

Recently it was shown that  quantum-gravitational effects may introduce 
deformations on the black hole surface, which, although complex and dynamical, 
as a  first approximation can  effectively and coarse-grained be described 
by a fractal structure. As a result, the black hole entropy will deviate from 
the standard Bekenstein-Hawking one, given by the expression 
\cite{Barrow:2020tzx}
\begin{equation}
\label{Barrsent}
S_B=  \left (\frac{A}{A_0} \right )^{1+\Delta/2}, 
\end{equation}
with $A$ the standard black-hole area  and $A_0$ the Planck area. Hence, 
the quantum-gravitational deformation is quantified through the 
exponent $\Delta$,  lying  between the extreme values 
$\Delta=0$, which corresponds to the   standard 
Bekenstein-Hawking entropy, and  $\Delta=1$, which 
corresponds  to   the most intricate and deformed structure. 

Later on, new developments have appeared in the literature aiming to 
test the performance of the above Barrow entropy in the cosmological framework. 
The validity and the constraints imposed by the generalized
second law of thermodynamics, including the matter-energy content 
and the horizon entropy, were investigated in \cite{Saridakis:2020cqq}. 
Additionally, the Barrow holographic dark energy model was proposed in 
 \cite{Saridakis:2020zol} and has been tested 
against the latest cosmological data in 
\cite{Anagnostopoulos:2020ctz,Dabrowski:2020atl}  where
it was found that it describes very efficiently the late accelerated expansion 
of the universe, having additionally the correct asymptotic behavior 
\cite{Mamon:2020spa}. Finally, Barrow entropy has been studied in the black-hole 
context in 
\cite{Abreu:2020cyv,McInnes:2020qqh,Barrow:2020coo,Abreu:2020rrh,Abreu:2020dyu, 
Abreu:2020wbz}.
 
 In \cite{Saridakis:2020lrg} Barrow entropy was applied in the framework of 
``gravity-thermodynamics'' conjecture 
\cite{Jacobson:1995ab,Padmanabhan:2003gd,Padmanabhan:2009vy}, according to which
the first law of thermodynamics can be   applied   on the universe apparent 
horizon. As a result, one obtains a modified cosmology, with  extra terms in 
the Friedmann equations depending on the new exponent $\Delta$, which  disappear 
in the case $\Delta=0$, i.e when
  Barrow entropy becomes the standard  Bekenstein-Hawking one. Although this 
construction is very efficient in describing the late-tile universe, one should 
carefully examine whether the aforementioned extra terms are sufficiently small 
in order not to spoil the early-time behavior and in particular the Big Bang 
Nucleosynthesis (BBN) epoch.

In the current article we address the above crucial issue concerning the 
behavior of modified cosmology through Barrow entropy in the early universe.
Since it is known that a given cosmological model is considered viable if and 
only if it satisfies the appropriate conditions imposed by 
 BBN, we can impose the BBN observational requirements in order to extract 
constraints on the exponent $\Delta$ of Barrow entropy.
  The structure of the article is as follows: In Section \ref{mobdel} we 
present the modified Friedmann equations that arise from the 
``gravity-thermodynamics'' application of Barrow entropy. In Section 
\ref{BarBNanal} we perform the investigation of the BBN epoch in such a 
cosmological scenario, and we extract the constraints on the Barrow exponent. 
Finally, Section \ref{Conclusion} contains a summary of our results.

\section{Modified cosmology through Barrow horizon entropy}
\label{mobdel}
 
In this section we briefly review the construction of  modified Friedmann 
equations through the   application of   ``gravity-thermodynamics'' 
conjecture using  Barrow entropy \cite{Saridakis:2020lrg}.  We consider   a
Friedmann-Robertson-Walker (FRW) metric
\begin{equation}
ds^2=-dt^2+a^2(t)\left(\frac{dr^2}{1-kr^2}+r^2d\Omega^2 \right),
\end{equation}
  with $a(t)$   the scale factor, and where $k=0,+1,-1$ 
corresponds  
to flat, close and open   geometry respectively.
 Moreover, we assume that the universe is  filled with 
matter and radiation perfect fluids.

Let us start by presenting the above procedure in the usual case of 
general relativity. 
The gravity-thermodynamics
conjecture states that the first law can be   applied   on the universe horizon 
 considered as a thermodynamical system 
separated  by a causality barrier
\cite{Jacobson:1995ab,Padmanabhan:2003gd,Padmanabhan:2009vy}, with the standard
choice being  the     apparent one 
\cite{Frolov:2002va,Cai:2005ra,Cai:2008gw}
\begin{equation}
\label{FRWapphor}
 {r_{A}}=\frac{1}{\sqrt{H^2+\frac{k}{a^2}}},
\end{equation}
with $H=\frac{\dot a}{a}$   the Hubble parameter and dots 
denoting time-derivatives.
For the horizon temperature one uses the corresponding  black hole  expression, 
but with the apparent horizon replacing the black-hole one 
\cite{Gibbons:1977mu}, namely 
\begin{equation}
\label{Th}
 T_h=\frac{1}{2\pi{r_{A}}}.
\end{equation}
The apparent horizon entropy will be given by the  black-hole one in a similar 
way, namely from   Bekenstein-Hawking relation $S=A/(4G)$, with $A=4\pi 
r_A^2$   its area   and $G$ the 
gravitational constant (in units where $\hbar=k_B = c 
= 1$):
\begin{equation}
\label{FRWHorentropy}
S_h=\frac{1}{4G} A.
\end{equation}
Hence, incorporating the energy flow through the horizon $\delta 
Q=-dE=A(\rho_m+p_m+\rho_r+p_r)H {r_A}dt$
with  $\rho_i$ and $p_i$ the energy density and pressure of the conserved 
matter and radiation fluids, and using the   first law of 
thermodynamics  $-dE=TdS$ as well as expressions 
(\ref{Th}),(\ref{FRWHorentropy}), we can finally extract the Friedmann 
equations \cite{Cai:2005ra}
\be
\label{FRWcFE1}
-4\pi G (\rho_m +p_m+\rho_r+p_r)= \dot{H} - \frac{k}{a^2}.
\ee 
\be 
\label{FRWcFE2}
\frac{8\pi G}{3}(\rho_m+\rho_r) =H^2+\frac{k}{a^2}-\frac{\Lambda}{3},
\ee
where the cosmological 
constant arises as an   integration constant.
Note that in the above steps, and in order to avoid non-equilibrium 
thermodynamics, one 
  applies the usual  equilibrium assumption, namely  that the   universe fluids 
have   the same temperature    with the
horizon 
\cite{Izquierdo:2005ku,Padmanabhan:2009vy,Frolov:2002va,Cai:2005ra,
Akbar:2006kj}.

 The gravity-thermodynamics conjecture has been widely and efficiently   
  applied    in many  modified theories of 
gravity, as long as one uses    the 
modified entropy relation corresponding to  each theory 
\cite{Akbar:2006er,Paranjape:2006ca,Sheykhi:2007zp,Jamil:2009eb,
Cai:2009ph,
Wang:2009zv,Fan:2014ala,Lymperis:2018iuz,Arias:2019zug}.  
Knowing the above, one can   apply   the   gravity-thermodynamics 
conjecture in the case of  Barrow entropy. Hence, 
using  (\ref{Barrsent}) instead of (\ref{FRWHorentropy})
 one finally results to  \cite{Saridakis:2020lrg}
\begin{equation}
 \label{FRWgfe1}
-\frac{(4\pi)^{(1\!-\!\Delta/2)}}{2(2+\Delta)}A_{0}^{(1\!+\!\Delta/2)}
(\rho_m+p_m+\rho_r+p_r)= 
\frac{\dot{H}-\frac{k}{a^2}}{\left(H^2\!+\!\frac{k}{
a^2}\right)^{\Delta/2}},  
\end{equation}
and through integration to
\begin{eqnarray} \label{FRWgfe2}
&&
\!\!\!\!\!\!\!\!\!
\frac{2\!+\!\Delta 
}{2\!-\!\Delta} \left(H^2\!+\!\frac{k}{a^2}\right) 
^{1\!-\!\Delta/2}
=
\frac{ (4\pi)^{(1\!-\!\Delta/2)}A_{0}^{(1\!+\!\Delta/2)} }{6} 
(\rho_m+\rho_r)\nonumber\\
&&
\ \ \ \ \ \ \ \ \ \ \ \ \ \ \ \ \ \ \ \ \ \ \ \ \ \  \ \  \
+\frac{{C}}{3} A_{0}^{(1+\Delta/2)},
\end{eqnarray}
where ${C}$  is the integration constant. 
As we observe, we have resulted to   modified Friedmann equations 
which include extra  terms comparing to general 
relativity. Restricting for simplicity  in the flat case  
$k=0$, we 
can re-write 
them  as
\begin{eqnarray}
\label{FRWFR1}
&&H^2=\frac{8\pi G}{3}\left(\rho_m+\rho_r+\rho_{DE}\right)\\
&&\dot{H}=-4\pi G \left(\rho_m+p_m+\rho_r+p_r+\rho_{DE}+p_{DE}\right),\ \ \ \ \ 
\ 
\label{FRWFR2}
\end{eqnarray}
where we have defined the energy density and pressure of the  effective dark 
energy 
sector as
 \begin{eqnarray}
&&
\!\!\!\!\!\!\!\!\!\!\!\!\!\!\!\!\!\!\!\!\!\!
\rho_{DE}=\frac{3}{8\pi G} 
\left\{ \frac{\Lambda}{3}+H^2\left[1-\frac{ \beta (\Delta+2)}{2-\Delta} 
H^{-\Delta}
\right]
\right\},
\label{FRWrhoDE1}
\end{eqnarray}
\begin{eqnarray}
&& \!\!\!\!\!\!\!\!\!\!\!\!\!\!\!\!\!\!\!\!
p_{DE}= -\frac{1}{8\pi G}\left\{
\Lambda
+2\dot{H}\left[1-\beta\left(1+\frac{\Delta}{2}\right) H^{-\Delta}
\right] \right.\nonumber\\
&&\left.  \ \ \ \  \ \ \ \  \ \ \,
+3H^2\left[1- \frac{\beta(2+\Delta)}{2-\Delta}H^{-\Delta}
\right]
\right\},
\label{FRWpDE1}
\end{eqnarray}
  with  $\beta\equiv \frac{4(4\pi)^{\Delta/2}G}{A_{0}^{1+\Delta/2}}$ a parameter 
with 
dimensions $[L^{-\Delta}]$ and
$ \Lambda  \equiv 4{C}G(4\pi)^{\Delta/2}$ a parameter with dimensions  
$[L^{-2}]$   
(in units where $\hbar=k_B = c 
= 1$). 
As expected,  for $\Delta=0$ (which implies that $\beta=1$)   the above
modified 
Friedmann equations   reduce to $\Lambda$CDM 
scenario.
 
 Finally, note that applying the first Friedmann equation (\ref{FRWFR1}) at 
present time   one obtains
\begin{equation} \label{lambdarad}
 \Lambda=\frac{3\beta 
(2+\Delta)}{2-\Delta}H^{(2-\Delta)}_{0}-3H^{2}_{0}\left(\Omega_{m0}+\Omega_{r0}
\right),
\end{equation}
 which is a convenient expression  relating $\Lambda$, 
$\Delta$ and $\beta$ with the present values of the matter and radiation 
density parameters $\Omega_{m0}$, an $\Omega_{r0}$, as well as with the present 
value of the Hubble parameter   $H_0$.

\section{Big Bang Nucleosynthesis  constraints on Barrow exponent $\Delta$}
\label{BarBNanal}

In this Section   we examine the Big Bang Nucleosynthesis (BBN) in the 
framework 
of modified cosmology through spacetime thermodynamics with Barrow entropy. 
Since BBN occurs in the radiation
  era we focus on the energy density of relativistic particles 
which is given by 
${\displaystyle \rho_r=\frac{\pi^2}{30}g_* {T}^4}$,
where the effective number of degrees of freedom is $g_*\sim 10$  
and ${T}$ is the temperature (the details of BBN are provided in 
the Appendix). The neutron abundance is 
calculated using the protons-neutron
conversion rate, i.e.
 \begin{equation}
 \lambda_{pn}({T})=\lambda_{(n+\nu_e\to p+e^-)}+\lambda_{(n+e^+\to p+{\bar
\nu}_e)}+\lambda_{(n\to p+e^- +
{\bar \nu}_e)}\,
 \end{equation}
and its inverse $\lambda_{np}({T})$, and the  total 
rate is therefore
 \begin{equation}\label{Lambda}
    \lambda_{tot}({T})=\lambda_{np}({T})+\lambda_{pn}({T})\,.
 \end{equation}
From (\ref{Lambda}) one can result to (see (\ref{LambdafinApp}) 
in the
Appendix)
 \begin{equation}\label{Lambdafin}
    \lambda_{tot}({T}) =4 A\, {T}^3(4! {T}^2+2\times 3! {\cal Q}{T}+2!
{\cal Q}^2)\,,
 \end{equation}
with ${\cal
Q}=m_n-m_p=1.29 \times10^{-3}$GeV   the  neutro-proton mass difference
and $A=1.02 \times 10^{-11}$GeV$^{-4}$.

Concerning the primordial mass
fraction of ${}
^4 He$, we  can   estimate it by   using   \cite{kolb}
 \begin{equation}\label{Yp}
    Y_p\equiv \lambda \, \frac{2 x(t_f)}{1+x(t_f)}\,,
 \end{equation}
with $\lambda=e^{-(t_n-t_f)/\tau}$, where $t_f$  is the freeze-out time 
of 
the weak
interactions, $t_n$   the corresponding  freeze-out time   of   
nucleosynthesis,
$\tau$ the neutron mean lifetime   (\ref{rateproc3}), and with
$x(t_f)=e^{-{\cal
Q}/{T}(t_f)}$   the neutron-to-proton equilibrium ratio.
The function $\lambda(t_f)$ accounts for the fraction of neutrons that 
decay into
protons inside the time interval $t\in [t_f, t_n]$.

 In any modified cosmological model one results with extra contributions in the 
Friedmann equations. The BBN happens at the 
radiation dominated era, and according to observations
these extra contributions 
need to be small comparing to the radiation sector in the concordance 
cosmological model, i.e the Standard Model radiation in the framework of 
General Relativity.
Hence,  the first Friedmann equation becomes 
approximately
\begin{eqnarray}
 H^2\approx\frac{8\pi G}{3} 
 \rho_r\equiv H_{GR}^2,
\end{eqnarray}
 and thus 
the scale factor evolves as 
$a\sim
t^{1/2}$, with $t$ the cosmic time.  The relation 
between temperature
and time  is therefore given by ${\displaystyle \frac{1}{t}\simeq 
\left(\frac{32\pi^3 
g_*}{90}\right)^{1/2}\frac{{T}^2}{M_{P}}
}$
(or
${T}(t)\simeq (t/\text{sec})^{-1/2} $MeV), which leads to 
\begin{eqnarray}
 H \approx    \left(\frac{4\pi^3 
g_*}{45}\right)^{1/2}\frac{{T}^2}{M_{P} 
},
\end{eqnarray}
   with $M_P=(8\pi G)^{-1}=1.22 \times 10^{19}$ GeV the Planck mass.

 Now, if the interaction rate $ \lambda_{tot}({T})$ given in  
(\ref{Lambdafin}) is $ \frac{1}{H}\ll\lambda_{tot}({T}) $, i.e. if the 
expansion time is much smaller than the interaction time then as usual we can 
safely consider that all processes are in thermal equilibrium 
\cite{kolb,bernstein}. On the other hand, if  $ 
\frac{1}{H}\gg\lambda_{tot}({T}) $ then particles decouple since they do not 
have the necessary time intervals to interact. Hence, the temperature at which 
particles decouple, namely the
freeze-out
temperature $T_f$, corresponds to the equality time, i.e. when $ 
 {H}=\lambda_{tot}({T}) $. Since $H \approx    \left(\frac{4\pi^3 
g_*}{45}\right)^{1/2}\frac{{T}^2}{M_{P} 
}$, while $ \lambda_{tot}({T})\approx q
T^5$, with  $q=4A4! \simeq 9.8\times 10^{-10}$GeV$^{-4}$, the above equality  
    provides the freeze-out temperature as 
     \begin{eqnarray}\label{Tfreeze}
    T_f  =    \left(\frac{4\pi^3 
g_*}{45M_P^2 q^2}\right)^{1/6} \sim 0.0006\, \text{GeV}.
 \end{eqnarray}

In the case of modified cosmology, the Hubble function $H$ will deviate from 
$H_{GR}$, and hence the    freeze-out temperature ${T}_f$ will also present a 
deviation $\delta {T}_f$ from the GR result (\ref{Tfreeze}).  This will in turn 
induce a deviation of the fractional 
mass $Y_p$, given by
 \begin{equation}\label{deltaYp}
    \delta
Y_p=Y_p\left[\left(1-\frac{Y_p}{2\lambda}\right)\ln\left(\frac{2\lambda}{Y_p}
-1\right)-\frac{2t_f}{\tau}\right]
    \frac{\delta {T}_f}{{T}_f}\,,
 \end{equation}
where we have imposed $\delta {T}(t_n)=0$ due to the fact that  ${T}_n$ is 
fixed by the 
deuterium
binding energy
\cite{Torres:1997sn,Lambiase:2005kb,Lambiase:2012fv,Lambiase:2011zz}.
The observational estimations of 
the mass
fraction $Y_p$ of baryons
converted to ${}^4 He$
during the BBN epoch  are
\cite{Coc:2003ce,Olive:1996zu,Izotov:1998mj,Fields:1998gv,Izotov:1999wa,
Kirkman:2003uv, Izotov:2003xn}
 \begin{equation}\label{Ypoohdges}
 Y_p=0.2476\,, \qquad |\delta Y_p| < 10^{-4}\,.
 \end{equation}
 Hence, inserting these into (\ref{deltaYp}) we extract   the upper 
bound on 
$\frac{\delta
{T}_f}{{T}_f}$, i.e.
 \begin{equation}
 \label{deltaT/Tbound}
    \left|\frac{\delta {T}_f}{{T}_f}\right| < 4.7 \times 10^{-4}\,,
 \end{equation}
 which is the allowed deviation from standard cosmology.

In the scenario at hand, the Barrow-entropy-related effective dark energy 
$\rho_{DE}$ given in 
(\ref{FRWrhoDE1}) is in principle present at the BBN times too. Hence, this 
should be small comparing to $\rho_r$, and thus it can be  
treated as a perturbation, while the matter sector can be neglected as usual.  
Therefore, the Hubble function arises from (\ref{FRWFR1}) as
  \begin{eqnarray}\label{H11}
    H&=&H_{GR} \sqrt{1+\frac{\rho_{DE}}{\rho_r}}=H_{GR}+\delta H ,
 \end{eqnarray} 
and hence
   \begin{eqnarray}\label{deltaH}
     \delta H&=&\left(\sqrt{1+\frac{\rho_{DE}}{\rho_r}}-1\right)H_{GR}\,.
 \end{eqnarray} 
 Thus, this deviation $\delta H$ from   standard $H_{GR}$ will lead to $\delta 
{T}_f$, and since as we mentioned $H_{GR}= \lambda_{tot} \approx q {T}^5$, we 
find 
\begin{equation}\label{H_T=Lambda}
  \left(\sqrt{1+\frac{\rho_{DE}}{\rho_r}}-1\right)H_{GR} = 5q {T}_f^4 \delta 
{T}_f\,,
  \end{equation}
  which since $\rho_{DE}\ll \rho_r$ becomes
  \begin{equation}\label{deltaTfcon11}
  \frac{\delta {T}_f}{{T}_f}\simeq \frac{\rho_{DE}}{\rho_r}\frac{H_{GR}}{10 
q
{T}_f^5}\,.
\end{equation}

We have now all the information to proceed to the 
  investigation of the BBN bounds  
  on the   parameter  $\Delta$ of  Barrow entropy.
These constraints will be extracted using  (\ref{deltaTfcon11}) and 
(\ref{FRWrhoDE1}).
Additionally, we will  use the   numerical values
 \begin{equation}
 \Omega_{m0}=0.3\,,\quad  \Omega_{r0}=0.000092\,,  \quad {H}_0=1.4\times 
10^{-42}\,\mbox{GeV}\,.
 \end{equation}

Inserting (\ref{FRWrhoDE1}) into  (\ref{deltaTfcon11}), and eliminating 
$\Lambda$ using   (\ref{lambdarad}), we find 
\begin{eqnarray}\label{deltaTf}
  &&
\!\!\!\!\!\!\!\!\!\!\!\!\!\!\!
  \left|\frac{\delta {T}_f}{{T}_f}\right|=\! \left[g_* \pi^2 T_f^4 
(\Delta-2)\right]^{-1}\nonumber\\
&&\ \ \ \cdot
9 M_P^2\Big\{
2^{\frac{3\Delta}{2}} H_0^{  2\! - \!\Delta} M_P^\Delta \pi^\Delta (2 \!+ \!
\Delta)\! + \!
H_0^2 ( \Delta\!-\!2) \Omega_{m0} \nonumber\\
&&\ \ \
+ 
 q^2 T_f^{10} \!\left[2 \!-\! \Delta\! - \!
    2^{\frac{3\Delta}{2}} M_P^{\Delta} \pi^\Delta (q T_f^5)^{-\Delta} (2 \!+ \!
\Delta)\right] \!\!
\Big\}.
\end{eqnarray}
Since all constants are known, if we insert the above expression into 
(\ref{deltaT/Tbound}) we obtain the BBN bounds on Barrow $\Delta$. As expected 
for $\Delta=0$ we obtain $\delta {  T}_f/{  T}_f=0$.
 
\begin{figure}[ht]
\hspace{-0.8cm}
\includegraphics[width=8.7cm]{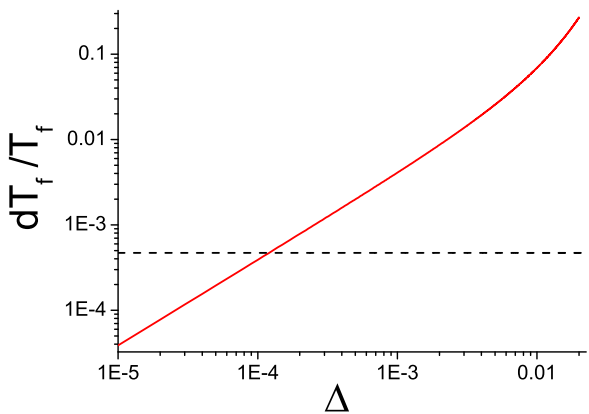}
\caption{\label{Const1} \textit{ $\delta {  T}_f/{  T}_f$ from
(\ref{deltaTf})
vs $\Delta$  (red solid  curve)   and 
the upper bound
for $\delta {  T}_f/{  T}_f$ from (\ref{deltaT/Tbound}) (black dashed line). 
As 
we can
see,
constraints from BBN require  $\Delta\lesssim 1.4\times 10^{-4}$.}}
\end{figure}

 In Fig. \ref{Const1} we depict $\delta {  T}_f/{  T}_f$ from
(\ref{deltaTf})
vs $\Delta$ (red curve), as well as the upper bound from 
(\ref{deltaT/Tbound}). 
As we can see,
constraints from BBN require $\Delta\lesssim 1.4\times 10^{-4}$.

From the above analysis we conclude that if Barrow entropy is indeed the case 
in Nature, which through gravity-thermodynamics conjecture would thus result to 
modified cosmology, then the Barrow exponent should be inside the bound  
$\Delta\lesssim 1.4\times 10^{-4}$ in order not to spoil the BBN epoch. This is 
the main result of the present work and shows that the deformation from 
standard  Bekenstein-Hawking expression should   be quite small as 
expected.

\section{Conclusions}
\label{Conclusion}

In this work we have used Big Bang Nucleosynthesis analysis and data in order 
to impose constraints on the exponent $\Delta$ of Barrow entropy. The latter is 
an extended entropy relation arising from the incorporation of 
quantum-gravitational effects on the black-hole structure, parameterized 
effectively by the new parameter $\Delta$. When considered in 
a cosmological framework and under the light of the gravity-thermodynamics 
conjecture, Barrow entropy leads to modified cosmological scenarios whose 
Friedmann equations contain extra terms.  This 
construction is very efficient in describing the late-time universe, 
nevertheless one should   examine whether the involved extra 
terms are sufficiently small in order not to spoil the early-time behavior and 
in particular  BBN  epoch.

We performed a detailed analysis of the BBN era in the above new cosmological 
scenarios  and we calculated the deviation of the freeze-out
temperature comparing to the result of standard cosmology, brought about by 
Barrow entropy exponent $\Delta$. Hence, we used the observationally determined 
bound on $ \left|\frac{\delta {T}_f}{{T}_f}\right|$ in order to extract the 
upper bound on $\Delta$. As we showed, the Barrow exponent should be 
inside the bound  $\Delta\lesssim 1.4\times 10^{-4}$ in order not to spoil the 
BBN epoch. As expected the latter  result  shows that the deformation from 
standard  Bekenstein-Hawking expression should be small.

It would be interesting to investigate the case where the complexity and 
dynamicallity of quantum-gravitationally deformed horizon structure would be 
incorporated through a Barrow exponent that depends on time and scale, as it 
has already been done with Tsallis entropy exponent \cite{Nojiri:2019skr}. This 
construction could leave more freedom in the deviation of Barrow entropy from  
Bekenstein-Hawking one. However, such a detailed study lies beyond the scope of 
the present work and it is left for a future project.

\appendix*

\section{Big Bang Nucleosynthesis}

In this Appendix we  review briefly the  Big Bang 
Nucleosynthesis features 
\cite{kolb,bernstein}.
The energy density of relativistic particles ($T\gg m, \mu$, where $\mu$ is the 
chemical
potential)
 filling up the early Universe  is   $\rho=\frac{g_s}{(2\pi)^2}\int E 
n(E/T)
d^3p=\frac{\pi^2}
{30}g T^4$, with $g_s$ denoting the degeneracy factors for particle species 
($g_\gamma=2$,
$g_{e}=4$, $g_\nu=2$), and $g=g_b+\frac{7}{8}g_f=\frac{43}{4}$ 
($g_f=g_e+3g_\nu=10$) are
the effective  
degrees of freedom (one assumes implicitly that muon and tau neutrinos have a 
small mass comparing to the effective temperature, and  that other massless 
species are not present).

The primordial ${}^4He$ in the   early Universe   was formed at a
temperature ${T}\sim
100$ MeV. The number and energy densities were formed by photons 
and relativistic leptons (electron,
positron and neutrinos), while rapid collisions were forcing all these 
particles to be in
thermal equilibrium. In particular,     protons and neutrons were 
maintained in thermal
equilibrium through their interactions with leptons:
 \begin{eqnarray}\label{proc1}
    \nu_e+n & \longleftrightarrow & p+e^- \\
    e^++n & \longleftrightarrow & p + {\bar \nu}_e \label{proc2} \\
    n& \longleftrightarrow & p+e^- + {\bar \nu}_e\,. \label{proc3}
 \end{eqnarray}

 One can calculate the neutron abundance through the
conversion rate of protons 
to
neutrons, denoted by $\lambda_{pn}({T})$, and its inverse rate denoted by 
$\lambda_{np}({T})$. 
Hence,  at suitably high temperature  the weak interaction rates 
read as
 \begin{equation}\label{LambdaA}
    \lambda_{tot}({T})=\lambda_{np}({T})+\lambda_{pn}({T})\,.
 \end{equation}
 Now,   $\lambda_{np}$ is given by the sum of the rates corresponding 
to the processes
(\ref{proc1})-(\ref{proc3}), i.e.
 \begin{equation}\label{sumprocess}
    \lambda_{np}=\lambda_{(n+\nu_e\to p+e^-)}+\lambda_{(n+e^+\to p+{\bar
\nu}_e)}+\lambda_{(n\to p+e^- +
{\bar \nu}_e)},
 \end{equation}
while the rate $\lambda_{np}$ is obtained from  $\lambda_{pn}$ through
$\lambda_{np}({T})=e^{-{\cal Q}/{T}}\lambda_{pn}({T})$, where 
${\cal
Q}=m_n-m_p=1.29 \times10^{-3}$GeV is the neutron-proton mass difference.

During the freeze-out regime one can assume that
\cite{bernstein}: 
(i) The particles temperatures   are the same, namely
${T}_\nu={T}_e={T}_\gamma={T}$,
(ii) the temperature
${T}$ is lower from the typical energies $E$ that enter into the
integrals that appear in the expressions for the rates (and thus one can use 
the Boltzmann distribution $n\simeq e^{-E/{T}}$ instead of the
Fermi-Dirac one), (iii) $m_e\ll E_e, E_\nu$  i.e. the electron mass $m_e$ can 
be neglected comparing to the
electron and neutrino energies.

 From the above we conclude that  the interaction rate
of the process (\ref{proc1}) is \cite{kolb,bernstein} 
  \begin{equation}\label{rateproc1}
    d\lambda_{(n+\nu_e\to p+e^-)}= d\mu \,(2\pi)^4 |\langle{\cal M}|^2\rangle W 
\,,
  \end{equation}
with
\begin{eqnarray}
  d\mu & \equiv &  \frac{d^3p_e}{(2\pi)^3 2E_e} \frac{d^3p_{\nu_e}}{(2\pi)^3
2E_{\nu_e}}\frac{d^3p_
p}{(2\pi)^3 2E_p}\,, \label{dmu} \\
 {\cal M} &= &\left(\frac{g_w}{8M_W}\right)^2 [{\bar u}_p\Omega^\mu u_n][{\bar
u}_e\Sigma_\mu v_{\nu_e}]\,, \label{M} \\
  \Omega^\mu &\equiv & \gamma^\mu(c_V-c_A \gamma^5)\,,
  \\
  \Sigma^\mu&
  \equiv&
\gamma^\mu(1-\gamma^5)\,,\\
  W &\equiv & \delta^{(4)}({\cal P})n(E_{\nu_e})[1-n(E_e)]\,, \label{WA}\\
   {\cal P}  &   \equiv &  p_n+p_{\nu_e}-p_p-p_e  
\,.
 \end{eqnarray}
  Note that in (\ref{M}) we have made use of the condition $q^2 \ll M_W^2$, 
with $M_W$     
the $W$
vector gauge boson  mass, and where $q^\mu=p_n^\mu-p_p^\mu$  is the 
momentum transferred. From   (\ref{rateproc1}) we obtain
\begin{equation}\label{rateproc1fin}
    \lambda_{(n+\nu_e\to p+e^-)}=A \, {T}^5 I_y\,,
 \end{equation}
with
$
  A\equiv \frac{g_V+3g_A}{2\pi^3}\approx1.02 \times
10^{-11}$GeV$^{-4}$ \cite{bernstein},
and with
\begin{equation}
 I_y=\int_y^\infty \epsilon(\epsilon-{\cal Q}')^2\sqrt{\epsilon^2-y^2}\, 
n(\epsilon-{\cal
Q})[1-n(\epsilon)]d\epsilon,
 \end{equation}
 having defined
 \begin{equation}
 y\equiv \frac{m_e}{{T}}\,, \quad {\cal Q}'=\frac{{\cal Q}}{{T}}\,.
 \end{equation}

Repeating the above calculation steps for the process (\ref{proc2}) we acquire
  \begin{equation}\label{rateproc2fin}
    \lambda_{(e^+ + n \to p+ {\bar \nu}_e)}=A \, {T}^5 J_y\,,
 \end{equation}
 where
 \begin{equation}
 J_y=\int_y^\infty \epsilon(\epsilon+{\cal Q}')^2\sqrt{\epsilon^2-y^2}\,
n(\epsilon)[1-n(\epsilon+{\cal Q}')]d\epsilon\,,
 \end{equation}
and finally we extract
 \begin{equation}
 \label{ne-pnu-fin}
    \lambda_{(e^+ + n\to p+{\bar \nu}_e)}=A\, {T}^3(4! {T}^2+2\times 3! 
{\cal
Q}{T}+2! {\cal
Q}^2)\,.
 \end{equation}

Concerning  the neutron decay (\ref{proc3}) one has
 \begin{equation}\label{rateproc3}
    \tau=\lambda_{(n\to p+e^- +{\bar \nu}_e)}^{-1}\simeq 887 \text{sec}\,.
 \end{equation}
 As a result, in the incorporation  of the process (\ref{sumprocess}) we can 
safely neglect the neutron decay, namely the neutron can be
handled as a stable particle
during the Big Bang Nucleosynthesis.

In summary, the aforementioned approximations (i)-(iii) result to
\cite{bernstein}
 \begin{equation}
 \label{auxilirel}
  \lambda_{(e^+ +n\to p+{\bar \nu}_e)}=\lambda_{(n+\nu_e\to p+e^-)}\,.
 \end{equation}
 Hence, substituting   (\ref{auxilirel}) into (\ref{sumprocess}) and then
into (\ref{LambdaA}), leads to the expression for
$\lambda_{tot}({T})$ as
 \begin{equation}\label{LambdafinA}
    \lambda_{tot}({T})\simeq 2\lambda_{np}=4\lambda_{(e^+ +n\to p+{\bar 
\nu}_e)}\,,
 \end{equation}
 and hence inserting (\ref{ne-pnu-fin}) results to
  \begin{equation}\label{LambdafinApp}
    \lambda_{tot}({T}) =4 A\, {T}^3(4! {T}^2+2\times 3! {\cal Q}{T}+2!
{\cal Q}^2)\,.
 \end{equation}


\providecommand{\href}[2]{#2}\begingroup\raggedright
\end{document}